# Investigation on the impact of solar flares on the Martian atmospheric emissions in the dayside near-terminator region: Case Studies

L. Ram[1], R. Sharma[1], D. Rout[3], R. Rathi[1], S. Sarkhel[1,2*]

[*]Sumanta Sarkhel, Department of Physics, Indian Institute of Technology Roorkee, Roorkee - 247667, Uttarakhand, India (sarkhel@ph.iitr.ac.in)

[1]Department of Physics,
Indian Institute of Technology Roorkee,
Roorkee - 247667
Uttarakhand, India

[2]Centre for Space Science and Technology,
Indian Institute of Technology Roorkee,
Roorkee - 247667
Uttarakhand, India

[3]National Atmospheric Research Laboratory,
Gadanki, India

**Abstract**

Solar transient events like flares can cause sudden changes in planetary plasma and neutral environment. Here, we present an investigation of the variability of the Martian atmospheric emissions viz. OI 130.4 nm, 135.6 nm, $CO_2^+$ ultraviolet doublet (UVD), and CO Cameron band (CB) in the less explored dayside near-terminator region during solar flare events. The two X8.2 and M6 class flares during September 2017 on Mars have been selected from existing catalogs. Using data from the imaging ultraviolet spectrograph (IUVS) aboard the MAVEN spacecraft, we examined limb radiance profiles. We observed a significant increase in radiance for major emissions around the peak with a more pronounced impact below the peak during flares compared to quiet time. During solar flares, for 130.4 nm and 135.6 nm emission, the maximum deviation in radiance beneath peak approaches to ~63% and ~123%, respectively. Whereas, for $CO_2^+$ UVD and CO CB, it is ~64% and ~50%, respectively. Additionally, we have presented an average scenario of dayside near-terminator (SZA~70-90°) and observed a notable trend of higher percentage deviation for atomic emissions compared to molecular emissions during flares. Further, our analysis depicts a higher percentage deviation during X8.2 compared to M6 class flare. This study underscores that during flares, higher photoelectron impact and irradiance flux drive the production processes, leading to enhanced emissions. The case studies emphasize, for the first time, the significant influence of flares on the Martian dayside near-terminator region, advancing our deeper understanding the impact of varying solar flare intensities to the planetary atmosphere.

**Key Points**

1. During solar flares, increase in radiance of Martian atmospheric emissions occurs at peak altitudes with most pronounced deviation below it.

2. During the X8.2 flare, the maximum percentage deviation in peak radiance is observed to be greater in magnitude compared to the M6 flare.

3. The percentage deviation of flare mean peak radiance is observed to be larger for the atomic emissions compared to the molecular emissions.

**Plain Language Summary**

Sun emits photon radiation which arrives on Mars in nearly 12-13 minutes. With varying Sun's activity, the sudden changes in radiation especially X-ray and EUV fluxes during flares cause sudden disturbances in the Martian environment and atmospheric emissions. This motivates us to study the Martian atmospheric response with respect to varying solar flare intensities. In the present work, we focused on the major Martian atmospheric emissions i.e., OI 130.4 nm, 135.6 nm, $CO_2^+$ UVD, and CO Cameron band. This study, for the first time, highlights the variation of Martian dayglow in the dayside near-terminator region during two solar flares of different intensities. We have found that during the flares, there is a significant enhancement in the peak radiance (brightness) for all emissions at peak with a more pronounced effect at lower altitudes (~90-120 km) below the peak region. This is attributed to increased solar and photoelectron flux in the ionosphere during flares. In addition, we have also found that during flares, variation in mean peak radiance for atomic emissions is greater in comparison to molecular emissions in the dayside near-terminator. Therefore, these results shed light on the Sun-Mars interaction and Mars's response to the weak and strong flare over time.

# 1. Introduction

The Sun releases its energy in the form of electromagnetic radiation, charged particles, and magnetic fields. The origin of most of the space weather events e.g., solar flares, interplanetary coronal mass ejections (ICMEs), corotating (stream) interaction regions (CIRs/SIRs), and solar energetic particles (SEPs) events are directly driven by the Sun's condition. These solar events comprise the ejection of solar wind particles, magnetic fields, and radiations. Among these solar events, solar flare is one of the major energetic phenomena, caused by the release of magnetic energy stored near sunspots in the solar atmosphere through magnetic reconnection (Parker, 1979). The solar flares emit radiation across the entire electromagnetic spectrum (pronounced in X-ray and extreme ultraviolet (EUV)) with energies of the order of $\sim 10^{29}$-$10^{32}$ erg (Fletcher et al., 2011; Shibata, 2015). The flares are classified into different categories such as X, M, C, B, and A classes according to their irradiance (incident power per unit area) (Veronig et al., 2002). The X class corresponds to the largest GOES flux of more than $10^{-4}$ W m$^{-2}$ (Fletcher et al., 2011). The M-class flare has one order less irradiance flux compared to X class flare. These solar flare events pose a significant effect on the entire solar system, especially the planetary atmosphere, like Earth and Mars (Chamberlin et al., 2018; Fang et al., 2019).

Earlier studies reported the impact of space weather phenomena on the Martian atmosphere (Kajdič et al. 2021; Mayyasi et al. 2018; Thampi et al. 2021 and references therein). During the passage of ICMEs and CIRs, the topside ionosphere gets compressed, which leads to plasma density depletion with enhanced temperature and loss rate of heavy ions (Futaana et al., 2008; Jakosky et al. 2015b; Krishnaprasad et al., 2021; Ram et al., 2023, 2024; Thampi et al., 2018). Further, during the solar flares, an enhancement in ionization, heating (Elrod et al., 2018; Thiemann et al., 2015), total electron content (TEC) and increase in the plasma density have been observed using radio occultation (Haider et al., 2009, 2012; Mahajan et al., 2009) and in-situ observations (Cramer et al., 2020; Fallows et al., 2015; Lee Y. et al., 2018; Lollo et al., 2012; Mendillo et al., 2006; Thiemann et al., 2018). In addition, the study by Cramer & Withers (2023) found an expansion of the thermosphere, which led to increased plasma density magnitude at fixed altitudes and pressures during the flare of 10 September 2017. Moreover, using modeling, a few studies also reported an increment in the photochemical escape rate of various neutral and ionic species during space weather events (Cramer et al., 2020; Lee C. O. et al., 2018; Mayyasi et al., 2018). There are numerous studies in the past, which studied the Martian dayglow using multi-mission UV spectrograph (Barth et al., 1971; Chaufray et al., 2015; Jain et al., 2015; Stewart, 1972). Several observations of the Martian atmospheric emission profiles were used to retrieve the density and temperature profiles (Chaufray et al., 2009, 2015; Qin et al., 2020). In the study by Jain et al. (2018), an increase in brightness of $CO_2^+$ ultraviolet doublet (UVD) and CO Cameron band emissions was observed at 90 km altitude during the X8.2 class (10 September 2017) flare. Furthermore, using the solar flux models viz. SOLAR2000 (S2K) (Tobiska, 2004) and EUVAC (Richards et al., 1994), Jain & Bhardwaj (2012) demonstrated the impact of varying solar EUV flux. They found that the altitude of peak limb brightness is independent of solar EUV flux changes. Therefore, the previous studies using in-situ, UV spectrograph and occultation measurements have focused on analyzing the Martian compositional variation primarily at altitudes above 150 km during solar flares. However, the influence of solar flares at altitudes below 150 km is not assessed so far. At these altitudes (100-150 km), the thermosphere of Mars which is mainly composed of $CO_2$ and its photochemical product (Fox et al., 1996) in interaction with solar irradiance leads to peak brightness. Therefore, it is imperative to investigate the prominent UV emissions to understand the behavior of the primary atmosphere during flares. In addition, to the best of our knowledge, there is a lack of understanding of the less explored dayside near-terminator atmosphere. This motivates us to investigate the less explored dayside near-terminator intermediate region, that coupled strongly to both the lower and upper atmosphere (Bougher et al., 1999) and will enhance our understanding of the deposition of the energy budget, dynamics, and heating during the solar flares.

In this study, we have presented the behavior of major atmospheric limb profiles during two solar flare events of X8.2 and M6 class on 10 and 17 September 2017, respectively. In order to address the scientific questions mentioned earlier, we have made use of Mars's atmosphere and volatile evolution (MAVEN; Jakosky, 2015a) multi-instrument datasets. Albeit, it is worth noting that the in-situ measurements onboard MAVEN are limited to an altitude range of ~150-160 km above the Mars surface. Therefore, to monitor the primary thermospheric intermediate region below 150 km, we have harnessed the capabilities of the remote sensing instrument i.e., imaging ultraviolet spectrograph (IUVS; McClintock et al., 2015) onboard MAVEN. This study makes the first instance of examining the flare-induced effects on the major atmospheric emissions such as atomic oxygen (OI 130.4 and 135.6 nm), $CO_2^+$ UVD, and CO Cameron band emissions at Mars. Moreover, our investigation is focused on the dayside near-terminator (60° < solar zenith angle (SZA) < 90°) atmosphere, where various processes such as plasma transport, escape, and photochemical processes play a pivotal role in the atmospheric dynamics. Therefore, this research will deepen our understanding of how strong and weak flares influence Martian atmospheric emissions in the dayside near-terminator region. This paper is structured as follows: Section 2 provides details about the instrument datasets and their corresponding processing levels, Section 3 presents the results and observations, Section 4 provides a discussion about the variation of peak radiance associated with different emissions, and Section 5 offers concluding remarks for this study.

**2. Data**

We utilized the data from multiple instruments aboard the MAVEN spacecraft for this work. The MAVEN spacecraft was injected into the intended Martian science orbit in September 2014, with an apoapsis of ~6200 km and a periapsis of ~150-160 km. The extreme ultraviolet monitor (EUVM; Eparvier, 2015), which is a part of the Langmuir probe and waves (LPW; Andersson et al., 2015) aboard MAVEN is utilized to measure solar irradiance. It measures the solar irradiance in three broad wavelength bands (0.1-0.7 nm, 17-22 nm, and 121.6 nm). The MAVEN EUVM Level 2 calibrated irradiance datasets are used during this work. Further, the MAVEN EUVM Level 3, minutes averaged modelled irradiance spectra are also utilized. To measure the energy and angular distributions of electrons in the Mars environment, we take advantage of the solar wind electron analyzer (SWEA; Mitchell et al., 2015) dataset. SWEA is a symmetric, hemispheric electrostatic analyzer, that measures the solar wind electrons and ionospheric photoelectrons of 3-4600 eV in the Mars environment. The in-situ key parameter Level 2 datasets of SWEA (Dunn, 2023; Mitchell et al., 2015) are used for photoelectron distributions. The EUVM data were accessed using the NASA Planetary Data System (PDS), whereas, SWEA datasets were accessed through the Python Data Analysis and Visualization tool (PyDIVIDE; MAVEN SDC et al., 2020) and the NASA PDS.

The limb radiance profiles of the Martian primary atmospheric emissions (OI 130.4 and 135.6 nm, $CO_2^+$ UVD 288-289 nm, and CO Cameron band 180-280 nm) are generated using the imaging ultraviolet spectrograph (IUVS; McClintock et al., 2015) instrument onboard the MAVEN spacecraft. IUVS is a remote-sensing package that operates in multiple modes like limb scans, echelle mode, disk scans, coronal scans, and stellar occultations. Only the limb scan mode datasets are made in use during the present work. IUVS takes the FUV (115-190 nm) and MUV (190-340 nm) channels to observe the spectra of the Martian upper atmosphere. The Level 1c processed data pipeline (Schneider et al., 2015) is utilized to generate the limb emissions features. Furthermore, we take the advantage of MAVEN science data center (SDC) to identify the solar flares at Mars. In this study, we have presented influence of two X8.2 and M6 class solar flare events on the Martian atmospheric emissions.

## 3. Results

The Martian atmosphere substantially alters in a short span of time (few minutes to hours) in response to large amount of energy deposition during solar flares (Mendillo et al., 2006; Liu et al., 2007; Lollo et al., 2012). In order to examine the impact of flares on the Martian emissions, we have presented a scenario for the primary atmospheric emissions (OI 130.4 and 135.6 nm, $CO_2^+$ UVD 288-289 nm, and CO Cameron band 180-280 nm) variation during the X8.2 (GOES XRS class) solar flare on **10 September 2017** and M6 (EUVM estimate) on **17 September 2017** solar flare events. The IUVS observations during this period lie primarily in the northern hemisphere (3-50°N), and are free from dust storms and strong crustal magnetic fields. The following subsections describe the variability of atmospheric emissions in the dayside near-terminator region (SZA~60°-90°) during two solar flare events.

### 3.1. Observations of solar flare at Mars using EUV Monitor onboard MAVEN

Fig. 1 depicts the time series evolution of the solar soft X-ray (0-7 nm) irradiance observed by EUVM onboard MAVEN during two significant flares on 10 and 17 September 2017. In Fig. 1, the y-axis represents the solar irradiance (Watt $m^{-2}$) and the x-axis represents the days of the September 2017. The surge in irradiance during 10 and 17 September 2017 flares, occurs at nearly 16:40 UT and 11:00 UT, respectively. The red-colored vertical dashed lines (Fig. 1) indicate the flare time orbits 5718 and 5755 observations by IUVS. Although, there are orbits 5719 and 5756 also lie in the disturb time (flare), however we have constrained our analysis near to peak flux i.e., orbits (5718 and 5755) to observe the maximum deviation from quiet time. In addition, the blue and green-colored vertical dashed lines represent the pre- and post-flare (quiet-time orbits 5717, 5724, 5727, 5754, 5757, and 5759) observations, respectively. These quiet-time profiles are used as a reference to evaluate the impact of flares during the disturbed (flare) time profiles. Due to the MAVEN spacecraft orbit trajectory constraints (Jakosky et al., 2015a) and the orientation of the IUVS observation (remote sensing) toward the terminator region of Mars (Jain et al. 2018), it is difficult for the simultaneous observations of the emissions during the peak irradiance of the flares. That is why during the two flares, IUVS observation time (shown in Fig. 2-7) does not precisely coincide with the peak time of the flare. However, the measurements lie in the disturbed hours or near to peak hours during the respective flares. Furthermore, in order to compare the flare and non-flare time behavior of the atmospheric emissions, we have selected those quiet time orbit profiles which exist at the same SZA range as for the flare time orbit profiles. Here, those post-flares orbit profiles are considered, when EUV irradiance returns to the background condition.

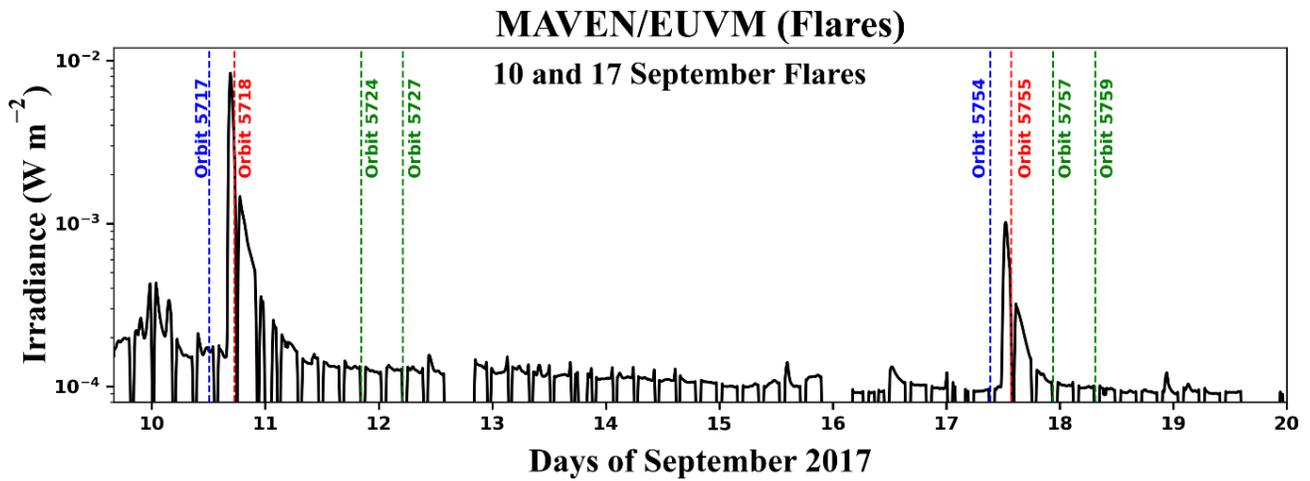

**Fig. 1.** Mars Atmosphere and Volatile EvolutioN/Extreme Ultraviolet Monitor (EUVM) solar irradiance in the 0.1-7 nm wavelength range during **10** and **17 September 2017** solar flare events at Mars. The x-axis represents the respective days of September. The red colored dashed lines (disturb orbits 5718 & 5755) represent flare time profiles whereas, the blue and green-colored vertical dashed lines represent the pre- and post-flare IUVS limb profiles (quiet orbits 5717, 5724, 5727, 5754, 5757 & 5759), respectively.

### 3.2. Impact on Martian atmospheric emissions in the dayside near-terminator during 10 and 17 September 2017 solar flares

We have presented the behavior of the OI 130.4 and 135.6 nm, $CO_2^+$ UVD, and CO Cameron band emissions. Fig. 2 shows the limb profiles of the atomic oxygen 130.4 nm and 135.6 nm emissions, which lie in the SZA between 66 and 90° during the 10 September 2017 solar flare at Mars. This flare is one of the most energetic flare events observed on Mars during the solar cycle 24 (Chamberlin et al., 2018; Theimann et al., 2018). The x-axis represents the radiance (kR, short for kilorayleigh), and the y-axis represents the tangent altitude between 80 and 170 km. A similar color scheme is used to represent the flare and non-flare (i.e., pre- and post-flare profiles) as indicated in Fig. 1. In Fig. 2, the solid and dashed colored profiles indicate the 130.4 nm and 135.6 nm emission, respectively. The flare, pre- and post-flare time limb profiles are selected in similar SZA for their suitable comparison (shown for each subplot in Fig. 2a-h). Although, IUVS gives 12 limb scans during a single MAVEN orbit, here we have shown only 8 limb scans (for flare events considered in this study) throughout the manuscript as four of the scans lie in the nightside.

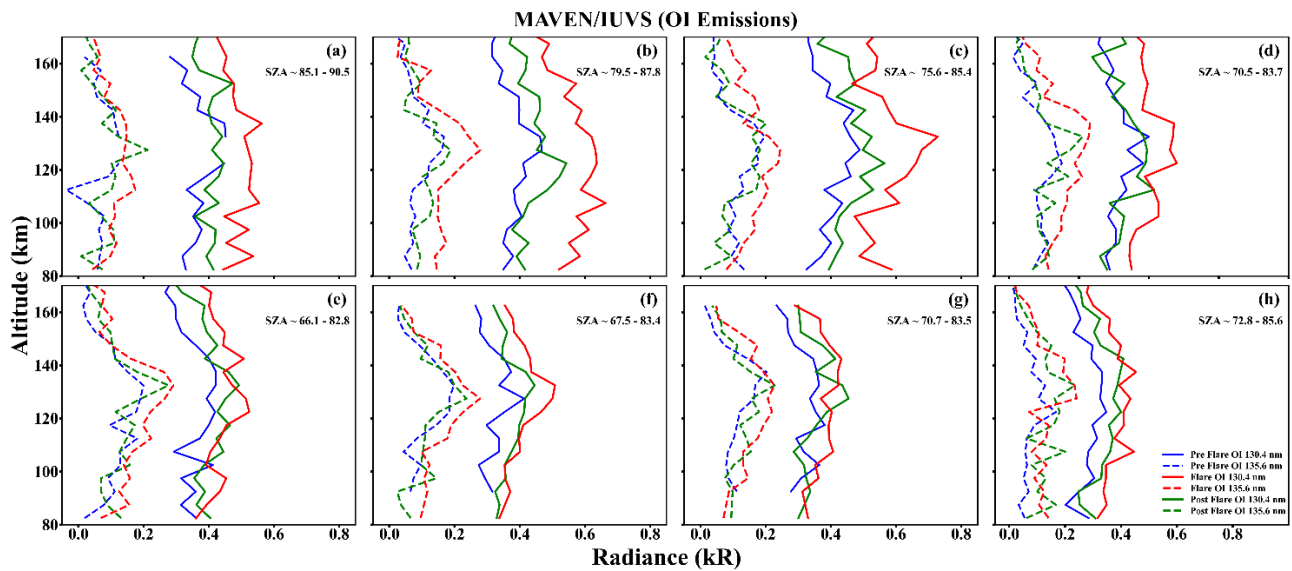

**Fig. 2.** Mars Atmosphere and Volatile EvolutioN/Imaging Ultraviolet Spectrograph (IUVS) far-ultraviolet 08 limb radiance profiles for OI 130.4 nm and OI 135.6 nm between 80 and 170 km tangent altitudes during **10 September 2017** solar flare at Mars. For 130.4 nm emission, the flare time profiles are shown in solid red, whereas pre- and post-flare profiles are represented by solid blue and green colors. Similarly, for 135.6 nm emission, all profiles are represented by dashed lines (same colors for quiet and disturb time profiles as for 130.4 nm). All profiles lie in the solar zenith angle (SZA) between 66 and 90°. The flare and non-flare profiles for each subplot are selected in a similar SZA range for better comparison.

During MAVEN orbit 5718 at around 17:40 UT (Fig. 1), the IUVS limb profiles for oxygen emissions (for both 130.4 and 135.6 nm) exhibited a significant increase in radiance at all tangent altitudes for each scan, lying in SZA range of 60-90°. However, the radiance profiles for SZA greater than 66° (Fig. 2 e, g) do not show a substantial change for both 130.4 and 135.6 nm emissions. After analyzing the limb radiance data for post flare orbit profiles, we have observed that during orbit 5727 (shown in green color, Fig. 1), the Martian atomic oxygen emission decreases and returned to background condition. However, the post-flare orbit profile radiance is near or slightly greater than pre-flare profiles. In order to calculate the deviation in radiance during flare from non-flare time, we have averaged the radiance during pre- and post-flare orbits to provide a quiet time condition. For 130.4 and 135.6 nm emissions, the maximum deviation in peak radiance during flare (shown in red color) from averaged quiet time magnitude is nearly 38% (solid line, Fig. 2c) and ~56% (dashed line, Fig. 2b), respectively. Also, during the flare, the calculated maximum peak ratio (130.4 nm/135.6 nm) is ~2.6. Furthermore, during the flare, the peak altitude of 130.4 nm (135.6 nm) exists between 132-142 km (127-132.5 km), whereas for the quiet time, it ranges between 127-132 km (127-137 km), respectively. In addition, at lower altitude below the peak, the maximum deviation of ~63% (~107 km altitude for 130.4 nm) and 123% (~92 km altitude for 135.6 nm) are observed during flare.

Fig. 3 presents the altitude radiance profiles of $CO_2^+$ UVD emission during the **10 September 2017** flare event for similar SZA mentioned above. The axes and color schemes are similar to Fig. 2. The limb profiles (MAVEN orbit 5718) show an enhancement in the peak radiance with most pronounced effect observed at lower altitudes at all SZA (66-90°). For this emission, we have observed that during orbit 5724 (shown in Fig. 1), the influence of flare gets reduced and atmosphere returned to its background condition. During the flare, the maximum deviation in $CO_2^+$ UVD peak radiance is nearly 13% from average quiet time magnitude (averaged peak magnitude of pre- and post-flare). Whereas, at lower altitude (~102.5 km altitude), the maximum deviation approaches to nearly 64% (Fig. 3e).

Similarly, Fig. 4 depicts the altitude radiance profile for CO Cameron band emissions. A similar trend has been observed for this emission as for $CO_2^+$ UVD. There is an enhancement in radiance at all SZA during flare time. During the flare, the maximum deviation in CO Cameron band peak radiance is nearly 14% from quiet time magnitude (averaged pre- and post-flare profiles). Whereas, at lower altitude (~102.5 km altitude), maximum deviation approaches to nearly 50% (Fig. 4d). Furthermore, during the flare, the peak altitude has not changed significantly in comparison to quiet time and exists between 122.5 and 132.5 km (similar for $CO_2^+$ UVD). It is evident from these observations that the highly energetic flare of 10 September 2017 had a significant effect on the Martian atmospheric emissions at peak with most pronounced impact observed at lower altitudes.

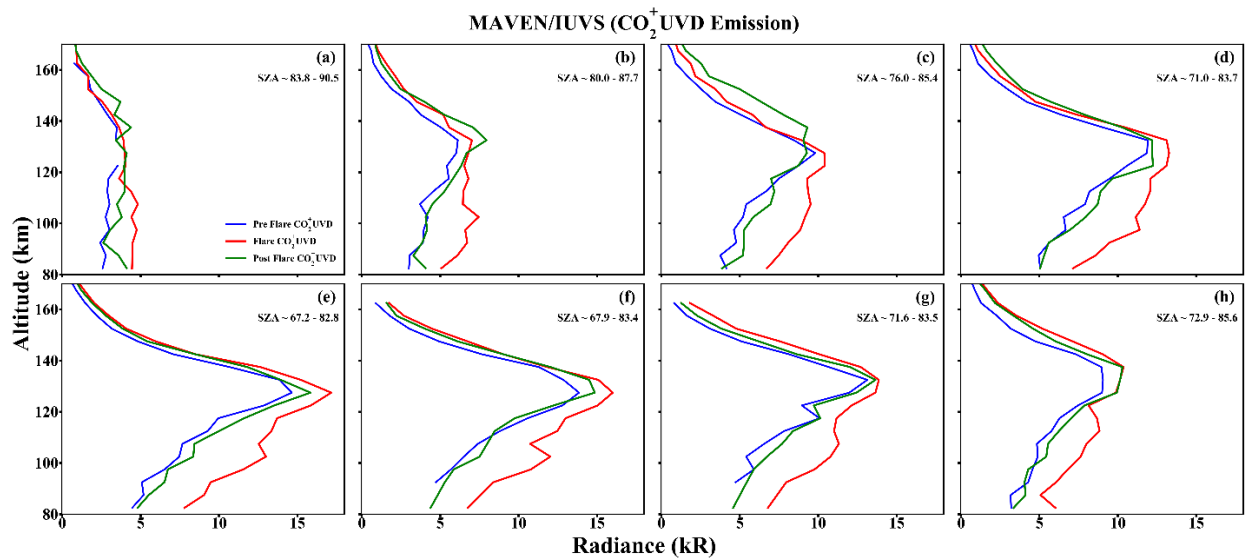

**Fig. 3.** Mars Atmosphere and Volatile EvolutioN/Imaging Ultraviolet Spectrograph (IUVS) mid-ultraviolet, 08 limb radiance profiles for $CO_2^+$ UVD (288-289 nm) between 80 and 170 km tangent altitudes during **10 September 2017** solar flare at Mars. The flare time profiles are shown in red-colored solid lines, whereas pre- and post-flare profiles are represented in blue and green colors, respectively. All profiles lie in the solar zenith angle (SZA) between 66° and 90°. The flare and non-flare profiles for each subplot are selected in a similar SZA range for better comparison.

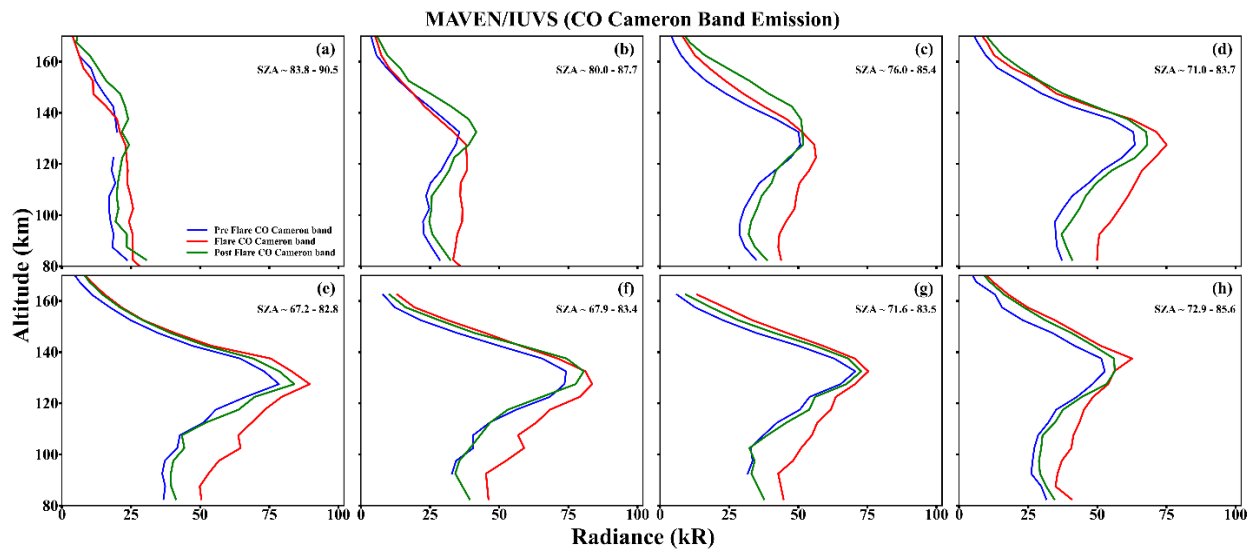

**Fig. 4.** Mars Atmosphere and Volatile EvolutioN/Imaging Ultraviolet Spectrograph (IUVS) mid-ultraviolet, 08 limb radiance profiles for CO Cameron (180-280 nm) band between 80 and 170 km tangent altitudes during **10 September 2017** solar flare at Mars. The flare time profiles are shown in red-colored solid lines, whereas pre- and post-flare profiles are represented in blue and green colors, respectively. All profiles lie in the solar zenith angle (SZA) between 66° and 90°. The flare and non-flare profiles for each subplot are selected in a similar SZA range for better comparison.

Fig. 5 depicts the limb profiles of the atomic oxygen 130.4 and 135.6 nm emissions over the near terminator (SZA ~60-83°) during the **17 September 2017** solar flare. The axes and the color schemes (for Fig. 5, 6 &7) for the flare and non-flare radiance profiles are similar to Fig. 2, 3 & 4. The flare profile is observed (MAVEN orbit 5755) nearly 13:35 UT (red color, Fig. 1). During the flare orbit 5755, the MAVEN IUVS observation is not directly coincided with the peak of flare (~12:30 UT, Fig. 1), but occur nearly one and a half hours later the peak irradiance. Further, we have observed that during orbit 5759 (shown in green color, Fig. 1), the increased Martian atomic oxygen emission (during flare orbit 5755) returned to background condition. During the flare (red color, Fig. 5), a significant increase in radiance is observed at all tangent altitudes which falls in the SZA range of 60-74°. However, the radiance profiles for SZA greater than 74° (Fig. 5 a, b) do not show a substantial change for both 130.4 and 135.6 nm emissions. The calculated maximum peak ratio (130.4 nm/135.6 nm) is ~2.23. The maximum magnitude of the peak radiance (red profile, Fig. 5g) for 130.4 nm emissions shows a deviation of nearly 30% from the quiet time peak radiance (averaged peak magnitude of pre- and post-flare). Whereas, for 135.6 nm emission (Fig. 5g), it is nearly 57% from quiet time, which is comparable to the change during 10 September for this particular emission. However, no notable change has been observed for peak altitude. It constrains between 122 and 132.5 km altitude for both quiet and disturbed time. Moreover, a more influence of flare is observed at lower altitudes below the peak in most of the subplots from pre-flare profile with lesser change from post-flare (for some of the subplots, shown in Fig. 5). The maximum deviation of 27% (~98 km altitude for 130.4 nm) and 105% (~98 km altitude for 135.6 nm) are observed during flare (Fig. 5g) from the quiet time.

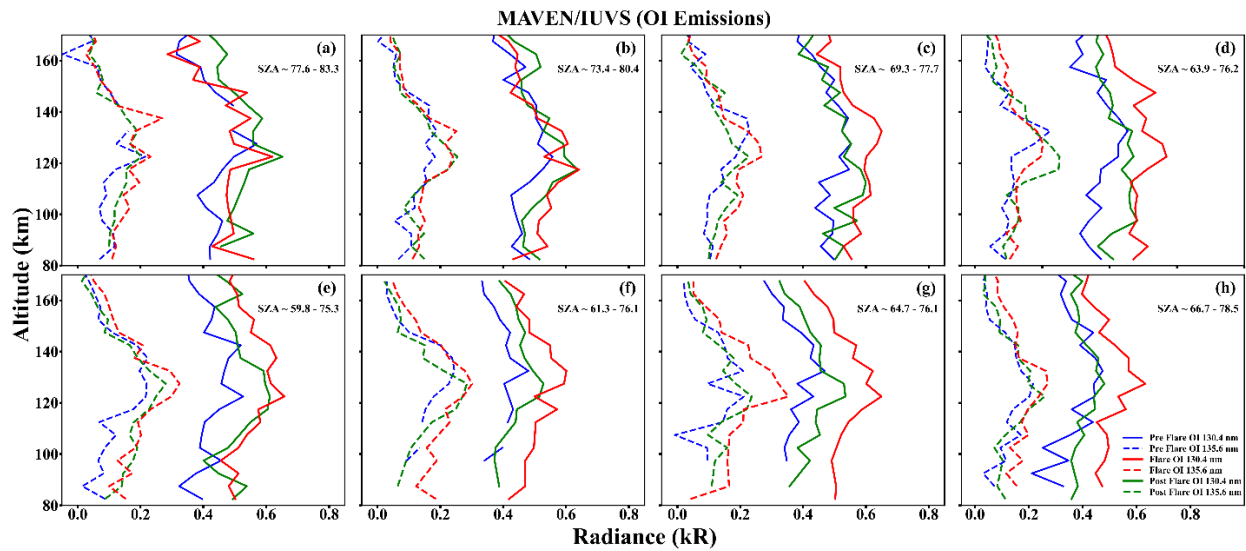

**Fig. 5.** Mars Atmosphere and Volatile EvolutioN/Imaging Ultraviolet Spectrograph (IUVS) far-ultraviolet 08 limb radiance profiles for OI 130.4 nm and OI 135.6 nm between 80 and 170 km tangent altitudes during **17 September 2017** solar flare at Mars. For 130.4 nm emission, the flare time profiles are shown in solid red, whereas pre- and post-flare profiles are represented by solid blue and green colors. Similarly, for 135.6 nm emission, all profiles are represented by dashed lines (same color for quiet and disturb time). All profiles lie in the solar zenith angle (SZA) between 60° and 83°. The flare and non-flare profiles for each subplot are selected in a similar SZA range for better comparison.

Fig. 6 depicts the altitude radiance profile for $CO_2^+$ UVD during the 17 September 2017 flare for similar SZA mentioned in the Figure 5. The $CO_2^+$ UVD emission shows an increase in radiance during orbit 5755 (flare time, shown in red color) compared to pre-flare profile (blue). The increased $CO_2^+$ UVD emission during flare (5755) is appears to decrease and reverted to its background level during orbit 5757 (highlighted in green color). During the flare (red color, Fig. 6), a significant increase in radiance is observed at peak and below the peak, for SZA range of 60-74° as shown in Fig. 6a-e. Whereas, there is no significant variation of the radiance above the peak. In addition, we observed no significant change near peak region in two subplots (Fig. 6f-h), although, it is enhanced at lower altitude region below the peak during the flare. During the flare, the maximum deviation in $CO_2^+$ UVD peak radiance is nearly 10% from average quiet time magnitude (averaged peak magnitude of pre- and post-flare). Whereas, at lower altitude (~107.5 km altitude), the maximum deviation approaches to nearly 22% (Fig. 6f).

Fig. 7 illustrates the altitude radiance profile for CO Cameron band emissions during 17 September 2017. A similar trend in radiance is observed for this emission as for $CO_2^+$ UVD (Fig. 7 a-f). However, we do not observe a significant change near peak as well as at lower altitude region below the peak for last two subplots (Fig. 7g-h). During flare, the maximum deviation in CO Cameron band peak radiance (Fig. 7f) is nearly 12% from average quiet time magnitude (averaged pre- and post-flare profiles). Whereas, at lower altitude (~102.5 km altitude), the maximum deviation in enhancement approaches to nearly 11% (Fig. 7f). Furthermore, during the flare, the peak altitude has not changed significantly in comparison to quiet time and exists between 122 and 132 km (similar for $CO_2^+$ UVD) as observed during 10 September 2017. Although, the 17 September measurement of IUVS does not exactly coincide with the irradiance peak (Fig. 1, orbit 5755), however, lies in the disturbed period (nearly one and a half hours later from the peak irradiance). We observed a substantial alteration at the peak radiance and altitude in the Martian atmospheric emissions. As a result, the lesser magnitude solar flares that occurred on 17 September 2017, exerted a notable influence on Mars emissions. Furthermore, our observations also indicates that the most prominent impact being observed during the highly intense flare of 10 September 2017 which led to enhancement in emission with higher percentage deviation.

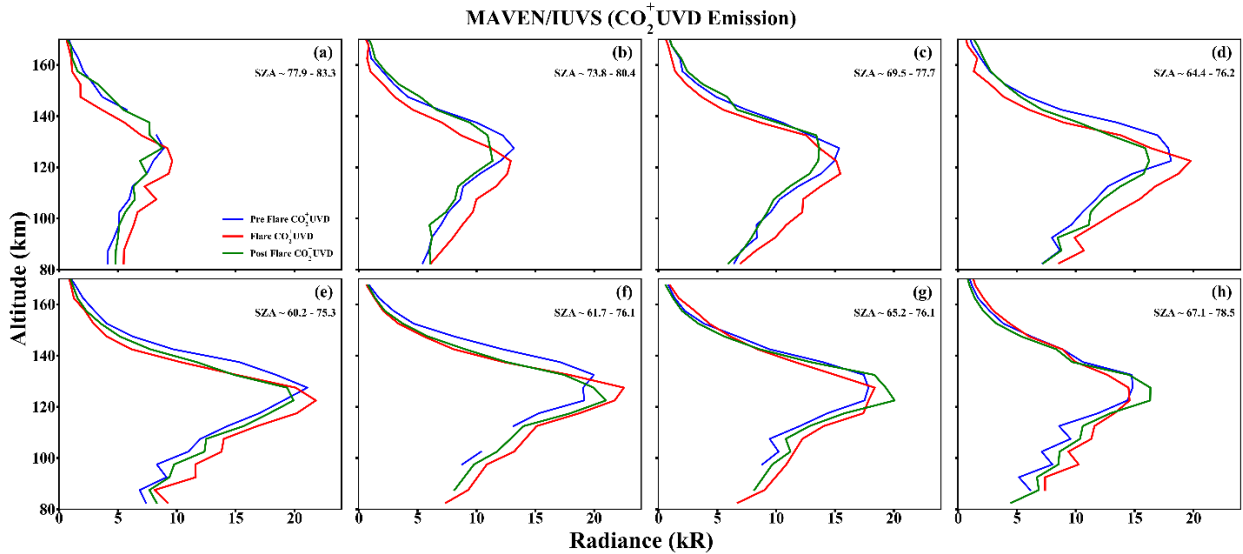

**Fig. 6.** Mars Atmosphere and Volatile EvolutioN/Imaging Ultraviolet Spectrograph (IUVS) mid-ultraviolet 08 limb radiance profiles for $CO_2^+$ UVD (288-289 nm) between 80 and 170 km tangent altitudes during **17 September 2017** solar flare at Mars. The flare time profiles are shown in red-colored solid lines, whereas pre- and post-flare profiles are represented in blue and green colors, respectively. All profiles lie in the solar zenith angle (SZA) between 60° and 83°. The flare and non-flare profiles for each subplot are selected in a similar SZA range for better comparison.

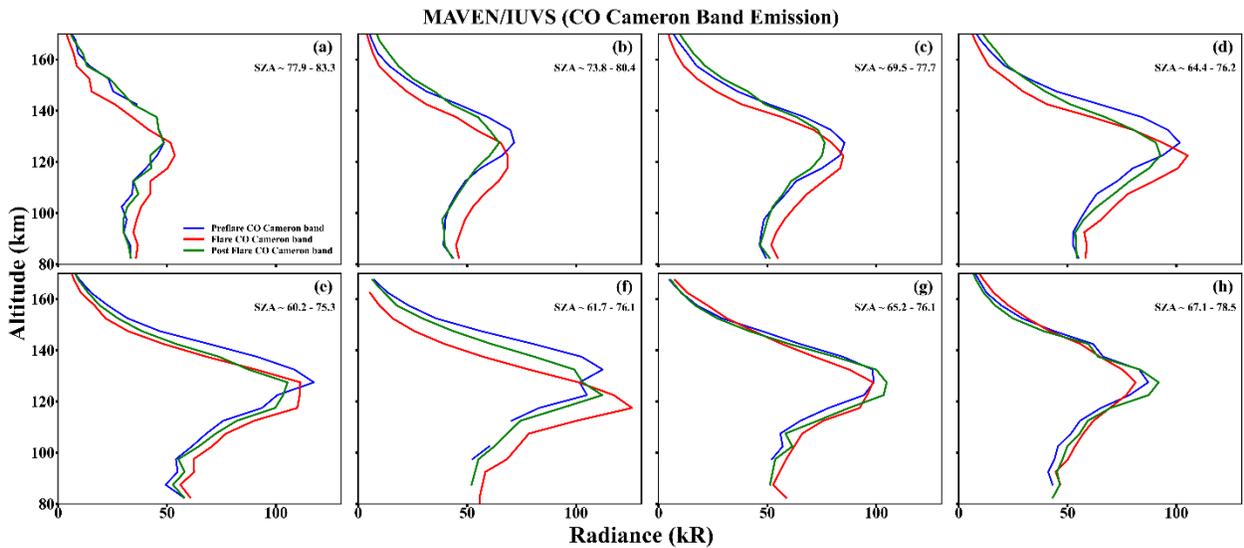

**Fig. 7.** Mars Atmosphere and Volatile EvolutioN/Imaging Ultraviolet Spectrograph (IUVS) mid-ultraviolet 08 limb radiance profiles for CO Cameron (180-280 nm) between 80 and 170 km tangent altitudes during **17 September 2017** solar flare at Mars. The flare time profiles are shown in red-colored solid lines, whereas pre- and post-flare profiles are represented in blue and green colors, respectively. All profiles lie in the solar zenith angle (SZA) between 60° and 83°. The flare and non-flare profiles for each subplot are selected in a similar SZA range for better comparison.

## 4. Discussion

This study is focused on the investigation of the Martian atmospheric emissions over northern latitudes (3-50° N) in the dayside near-terminator (SZA ~60-90°), during the occurrence of X8.2 (10 September 2017) and M6 (17 September 2017) class solar flares at Mars. The atmospheric limb radiance profiles for OI 130.4 and 135.6 nm wavelengths, $CO_2^+$ UVD, and CO Cameron band show a noticeable change during solar flares compared to non-flare periods within a similar SZA range. However, the peak altitude, where these emissions occurred, remained relatively stable during solar flares events. In addition, the atmosphere emissions for molecular species emission returned to normal condition in the early post-flare orbit (e.g., 5724, 5757) as compared to atomic O emissions, for which it appears to return to quiet condition in the post-flare orbits 5727, 5759 (Fig. 2-7). This could be due to more stability of atomic O compared to molecular $CO_2$ in the thermosphere (Shimazaki, 1989; Fox et al., 2017). The 130.4 nm (OI $^3S$ – $^3P$ transition) emission is primarily produced by resonance scattering and electron impact on O and $CO_2$ and can be given the following reactions (Ritter et al., 2019).

$$h\upsilon_{130.4} + O \rightarrow O(^3S) \tag{1}$$

$$e^- + O \rightarrow O(^3S) + e^- \tag{2}$$

$$e^- + CO_2 \rightarrow CO + O(^3S) + e^- \tag{3}$$

For the production of 135.6 nm (OI $^5S$ – $^3P$, a forbidden transition) emission, the possible sources are electron impact on O and electron dissociation of $CO_2$ (Mason, 1990; Ritter et al., 2019).

$$e^- + O \rightarrow O(^5S) + e^- \tag{4}$$

$$e^- + CO_2 \rightarrow CO + O(^5S) + e^- \tag{5}$$

However, Equation 5 has a minimal contribution of ~ 4-6% to 135.6 nm emission (Ritter et al., 2019). In addition, the $CO_2^+$ UVD is produced by the transition from the exited state $CO_2^+$ ($B^2\Sigma_u$) to the ground state $CO_2^+$ ($X^2\Pi$). The primary excitation mechanisms are electron impact and photoionization of $CO_2$ (Avakyan et al., 1998; González-Galindo et al., 2018; Gérard et al., 2019; Itikawa, 2002; Jain & Bhardwaj, 2012).

$$CO_2 + h\upsilon\ (\lambda < 69\ nm) \rightarrow CO_2^+\ (B^2\Sigma^+) + e^- \tag{6}$$

$$CO_2 + e^- \rightarrow CO_2^+\ (B^2\Sigma^+) + e^- + e^- \tag{7}$$

Resonance scattering of $CO_2$ is neglected due to its contribution less than 4% to $CO_2^+$ UVD emission (Fox and Dalgarno, 1979). Further, the CO Cameron band are produced by the forbidden transition from the exited state a³Π to ground state X¹Σ⁺ (Jain & Bhardwaj, 2012). The excitation mechanisms are the photodissociation of $CO_2$, electron impact dissociation of $CO_2$, electron impact excitation of CO, and the dissociative recombination of $CO_2^+$ (González-Galindo et al., 2018; Gérard et al., 2019; Johnson, 1972; Jongma et al., 1997).

$$CO_2 + h\upsilon \ (\lambda < 108 \text{ nm}) \rightarrow CO(a^3\Pi) + O \qquad (8)$$

$$CO_2 + e^- \rightarrow CO(a^3\Pi) + O + e^- \qquad (9)$$

$$CO + e^- \rightarrow CO(a^3\Pi) + e^- \qquad (10)$$

$$CO_2^+ + e^- \rightarrow CO(a^3\Pi) + O \qquad (11)$$

The electron impact (Equation 9) followed with Equations 10 and 11 are the dominant mechanism for CO (a³Π) (Gérard et al., 2019).

In the present study, we have found an enhancement in Martian atmospheric UV emissions during both flares, that characterized by a noticeable increase in the radiance around the peak region with the most pronounced impact observed at lower altitudes below the peak. However, the highly intense flare on 10 September 2017 exhibited a more efficient enhancement in the radiance profile in terms of magnitude. Fig. 8a illustrates variation in the irradiance flux during both flares (shown in solid and dashed red color lines for 10 and 17 September, respectively), as well as for non-flare time (shown by blue solid color line) using SWEA instrument onboard MAVEN spacecraft. The photoexcitation, photoionization, electron excitation, and electron impact process as mentioned in Equations 1 to 11 are the primary factors for the major atmospheric emissions (considered in this study). The high irradiance flux of X-ray (0.1-7 nm) and EUV flux (26-34 nm) as shown in Fig. 8a can drive the above-mentioned photoexcitation and photoionization processes (Equations 1, 6, and 8), which contribute to the higher magnitude of radiance as observed for each flare with most prominent impact on the 10 September flare. The study by Mendillo et al. (2006) found that increased fluxes over 0.1 to 5 nm range drive the changes in electron density at Martian lower altitudes (100-120 km) atmosphere. Further, the EUV fluxes (26-34 nm) were responsible for the enhancement around the peak (120-140 km). In the present study, we found an increase in the radiance at peak and beneath the peak region with higher percentage deviation during the flare of 10 September 2017. Whereas during the flare on 17 September 2017, the percentage deviation is lesser in terms of magnitude as compared to 10 September (below peak). This is because, during this flare, the measurement of the radiance profile (orbit 5755) occurred 1.5 hours after the peak

X-ray flux (Fig. 1) and coincided with the declining phase. In addition, the lower irradiance flux over 0.1 to 5 nm range and EUV (26-34 nm) flux (shown in dashed red color line in Fig. 8a), which is responsible for the enhancement at near peak and beneath the peak region, subsequently results in lesser change in magnitude compared to the 10 September 2017 flare.

Besides, the photochemistry-driven processes, the photoelectron impact processes (Equations 2 to 5, 7, 9 and 10) can also play a pivotal role in dictating the Martian atmospheric emissions (González-Galindo et al., 2018; Jain & Bhardwaj, 2012; Lee et al., 2021 and references therein). The direct excitation of the electron impact on oxygen during enhanced photoelectron flux results in the production of OI 130.4 nm & 135.6 nm emissions (Ajello et al., 2019; Zipf & Erdman, 1985), and $CO_2^+$ UVD (Avakyan et al., 1998; Itikawa, 2002) & CO Cameron band (Jain & Bhardwaj 2012; Johnson, 1972; Jongma et al., 1997; Lee et al., 2021). Fig. 8b depicts the integrated photoelectron flux (5-100 eV) variation during two flares observed below 500 km altitude. The significant enhancement in the photoelectron flux during the flare orbit (shown in vertical red-colored dotted lines). Also, the flux attains its maximum during 13 September 2017, which is primarily due to the passage of ICME at Mars after 10 September flare. However, our observations are constraining to flare time and are not lying under the influence of ICME. During flare orbits, the increasing electron flux subsequently results in escalation of the electron impact processes on major atmospheric species. Further, it generates energetic electrons in collisions with neutrals, resulting in higher magnitude of radiance. The studies by Mahajan et al. (2009) and Haider et al. (2011) demonstrated that there was an enhancement in the electron density during solar flares at all altitudes between 90 and 160 km compared to quiet time, which increased the electron impact processes. Furthermore, Mendillo et al. (2006) found that the flare-induced enhancement in electron density reached in the range of 50-200% compared to quiet time at lower altitudes (below peak region). The larger magnitude of the difference in radiance at the peak as well as at altitudes below the primary peak during both flares in the present study aligns with the enhanced electron density as mentioned in Mendillo et al. (2006) and Mahajan et al. (2009) (shown in Fig. 3, 4, 6, and 7). In addition, the higher magnitude of the solar irradiances at X-ray (0.1-7 nm) and EUV flux (shown in Fig. 8a) during both flares compared to the quiet time reinforces the enhancement in the electron density (Mendillo et al., 2006), resulting in higher chances of energetic electron impact processes on $CO_2$. This leads to higher emission near peak with more pronounced impact at lower altitudes. Therefore, the emissions are strongly affected by the X-ray irradiance, EUV flux as well as electron impact processes as mentioned above.

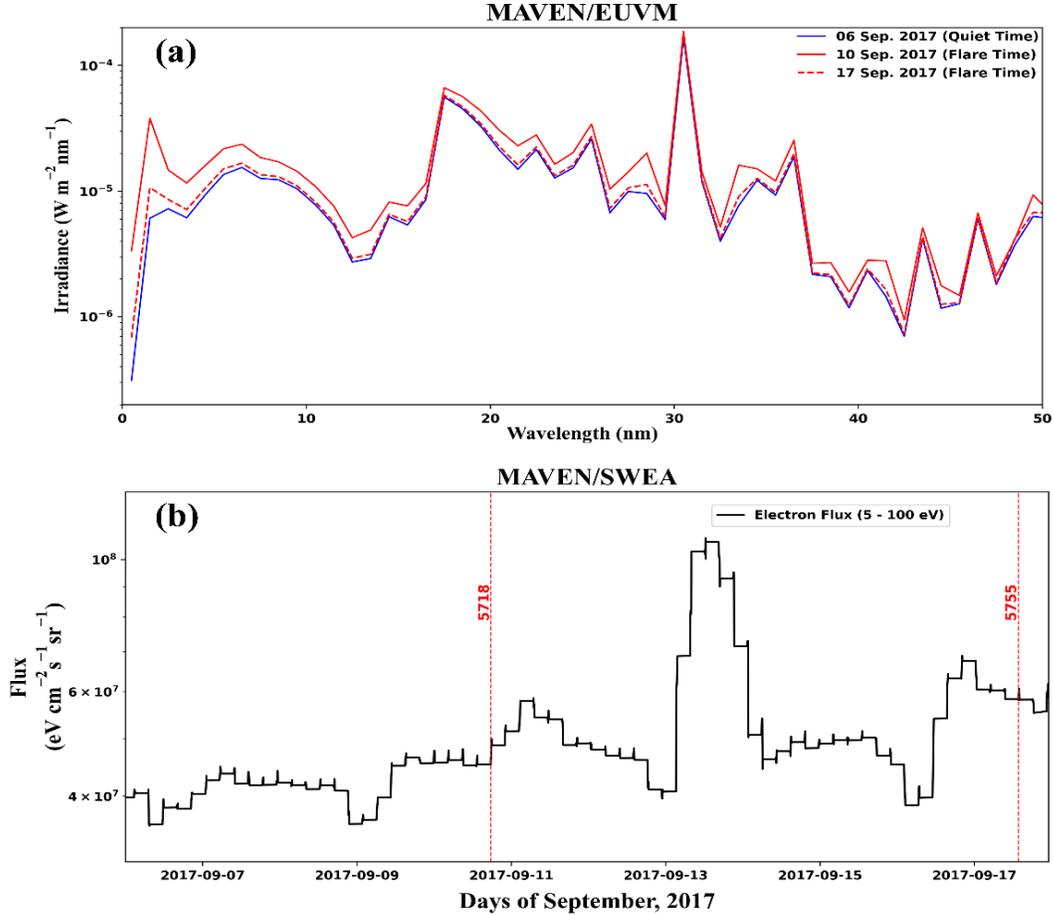

**Fig. 8. (a)** Extreme Ultraviolet Monitor (EUV) irradiance during the days of **06, 10, and 17 September 2017** represented by solid blue, solid red, and dashed red colors in a 1 nm bin for the wavelength between 0 and 50 nm. The blue profile denotes the non-flare, whereas the remaining color profiles represent the flare time irradiances (also show in legends). **(b)** Mars Atmosphere and Volatile EvolutioN/Solar Wind Electron Analyzer (SWEA) measurement of the photoelectron energy flux in the range of 5-100 eV during 06-17 September 2017 at Mars (below 500 km altitude). The vertical red-colored dashed lines represent the enhanced electron energy flux during flare time orbit profiles.

Furthermore, the observations during both flares were concentrated within the dayside near-terminator atmospheric region. In order to provide an average scenario of the dayside near-terminator, we calculated the mean peak radiance between 70 and 90° SZA. Table 1 provides a detailed information about the mean peak radiances with STDEV (standard deviation) as well as the percentage deviation during flares from the quiet time (averaged pre- and post-flare radiance profiles) for each major atmospheric emission. We observed a discernible trend from the analyses that during each flare, the percentage deviation in the mean peak radiance for atomic oxygen (OI 130.4 and 135.6 nm) exceeded that molecular species ($CO_2^+$ UVD, and CO Cameron band) emissions. Notably, the percentage deviation for each emission is significantly larger during intense X8.2 class flare in comparison to the M6 class flare considered in this study. In addition, the maximum deviation (Figures 2, 3, 4, 5, 6, and 7, see Section 3) at peak and below the peak is higher for atomic O compared to the molecular $CO_2$ and CO emissions during both flares. Therefore, the behavior of Martian emissions is in similar line with averaged behavior shown in Table 1. Moreover, higher deviation in OI 135.6 nm at lower altitudes (described in detail in Section 3) compared to 130.4 nm is primarily by the dependence of 135.6 nm on the electron impact process (Equations 4-5), whereas, 130.4 nm emission is dominated by resonant scattering (Equation 1) with a contribution of nearly 80% (Ritter et al., 2019). Therefore, enhanced photoelectron flux and irradiance (primarily X-ray and EUV, which enhance the electron density, Mendillo et al. 2006; Mahajan et al., 2009) could lead to the dominant role of photoelectron processes. This causes higher percentage deviation for 135.6 nm emissions (at peak and lower the peak) than 130.4 nm during solar flare event specifically, during intense X8.2 class flare event on 10 September 2017. Further, at lower altitude the $CO_2^+$ UVD emission is higher than CO Cameron band by 11-14% during both flares (shown in Section 3). This may be due the direct dependency of production processes (Equations 6-7) on the $CO_2$ local density (Gérard et al., 2019). Further, it was found that during flare orbit, $CO_2$ density quickly increased compared to its baseline values above peak altitude (Cramer et al., 2020) and decrease at lower altitude region (below peak) using derived $CO_2$ density (Jain et al., 2018, Figure 2 (right subplot)). Also, Jain et al. (2018) explained that the increase of $CO_2^+$ UVD emission due to increased SXR photons flux (see Fig. 8a of current manuscript) at lower altitude results in heating of atmosphere that led expansion with higher $CO_2$ density above peak (Cramer et al., 2020) during the 10 September 2017 solar flare event. Moreover, the study by Lee et al. (2022) found the lower cross-section magnitude for CO (~1.6 × $10^{-17}$ $cm^2$) compared to $CO_2$ (4.4 × $10^{-17}$ $cm^2$) at 20 eV photoelectron. For example, around 18.1 eV photoelectron is required to excite $CO_2^+$ ($B^2\Sigma^+$) (Equation 7). All these factors may increase the deviation for $CO_2^+$ UVD compared to CO Cameron band during the solar flare events. Although, for the averaged peak radiance behavior over near-terminator, the difference in percentage deviation for both molecular emissions i.e., $CO_2^+$ UVD and CO Cameron band lie in comparable range (Table 1).

|                      | 10 September 2017 (X8.2)                         |                                              |                              | 17 September 2017 (M6)                           |                                              |                              |
|----------------------|--------------------------------------------------|----------------------------------------------|------------------------------|--------------------------------------------------|----------------------------------------------|------------------------------|
| **Species Emissions** | **Mean Peak Radiance and STDEV (kR)** Non-Flare | **Mean Peak Radiance and STDEV (kR)** Flare | **Percentage Deviation (%)** | **Mean Peak Radiance and STDEV (kR)** Non-Flare | **Mean Peak Radiance and STDEV (kR)** Flare | **Percentage Deviation (%)** |
| **OI 130.4 nm**      | 0.4665 ± 0.064                                   | 0.573 ± 0.116                                | 23                           | 0.5526 ± 0.065                                   | 0.63861 ± 0.013                              | 15                           |
| **OI 135.6 nm**      | 0.1981 ± 0.036                                   | 0.2425 ± 0.040                               | 22                           | 0.2193 ± 0.023                                   | 0.2626 ± 0.011                               | 20                           |
| **$CO^+_2$ UVD**     | 10.662 ± 3.765                                   | 11.669 ± 4.212                               | 9.4                          | 11.871 ± 2.675                                   | 12.659 ± 2.932                               | 6.6                          |
| **CO Cameron band**  | 57.834 ± 19.762                                  | 63.335 ± 22.268                              | 9.5                          | 65.875 ± 14.996                                  | 69.028 ± 15.667                              | 5                            |

**Table 1:** The mean peak radiance along with Standard Deviation (STDEV) and percentage deviation (from quiet time), observed during periods of non-flare (quiet time) and flare (disturbed time) for emissions viz. OI 130.4 nm, OI 135.6 nm, $CO_2^+$ UVD, and CO Cameron band emissions in the Martian atmospheric dayside near-terminator region (SZA ~70-90°).

Our observations in the present study suggest the prominent role of photoelectron impact processes and solar irradiance flux in shaping the Martian emissions, alongside photochemistry. However, due to the unavailability of the MAVEN in-situ measurements below 150 km altitude, a comprehensive understanding of the thermosphere electron, ion, and neutral density cannot be fully utilized to explain the Martian emissions. Nevertheless, it provides an opportunity for upcoming Mars exploration missions to gather more sophisticated datasets of electron, ion, and neutrals to understand the dynamics of primary atmosphere (which exist below 150 km). Further, this would also motivate us to think about the local emission profile in addition to the current line-of-sight profile. Moreover, these case studies sheds light in our understanding of the Sun-Mars interaction, particularly during varying intensities of flares.

As this study includes only two cases of solar flare events and therefore, it cannot comprehensively conclude the inference that we made on the impact of flares on the Martian atmospheric emissions. In addition, determination of the dominant mechanisms during solar flares in proportion, remains unsolved. This work will be further improved by the inclusion of more numbers of such cases of X and M class flares and its impact on the Martian atmospheric species in order to build a significant statistics.

## 5. Summary and Conclusions

In this study, we investigated the impact of the solar flares during 10 (X8.2) and 17 (M6) September 2017 on the less explored Martian dayside near-terminator atmospheric region. The Martian major atmospheric emissions viz. atomic oxygen OI 130.4 and 135.6 nm, $CO_2^+$ UVD 288-289 nm, and CO Cameron band 180-280 nm have been investigated. Using MAVEN/IUVS remote sensing instrument datasets, we examined extensively the limb radiance profiles between 100 and 180 km altitudes at SZA ranging from 60-90°. Our research depicts that the limb radiance profiles for all emissions are strongly dependent on the strength of flares, resulting in augmented radiance at peak as well as at lower altitudes below the peak during flares compared to quiet time. During both X8.2 (M6) flares, the maximum deviation in peak radiance for OI 130.4 nm and 135.6 nm emissions is ~38% (~30%) and ~56% (~57%) with highest magnitude of ~63% (~27%) and ~123% (~105%), respectively at lower altitude region (below peak) from quiet time. Similarly, for $CO_2^+$ UVD and CO Cameron band, the peak deviation is ~13% (~10%) and ~14% (~12%) with highest magnitude of ~64% (~23%) and ~50% (~11%) at lower altitude, respectively. However, the peak altitude for each emission during flares is relatively stable. In addition, the present study also provides an average scenario of Martian emissions in the dayside near-terminator (SZA~ 70-90°). The mean peak radiance percentage deviation during both X8.2 (M6) solar flares for OI 130.4 nm and 135.6 nm emissions is ~23% (~15%) and ~22% (~20%), respectively. Similarly, for $CO_2^+$ UVD and CO Cameron band, the mean peak deviation is ~9.4% (~6.6%) and ~9.5% (~5%), respectively. The average analysis indicates that during both flares, the percentage deviation in the mean peak radiance for atomic oxygen (OI 130.4 and 135.6 nm) emission are greater in magnitude compared to the molecular ($CO_2^+$ UVD, and CO Cameron band) emission. These changes highlight the major role of photoelectron impact processes during flares along with photochemical processes. The increment in the solar flux primarily the X-ray (0.1-7 nm), EUV flux (26-34 nm), and photoelectron flux during both flares leads to enhancement in radiance around the peak. Therefore, the case studies considered in current work provides a new insight to understand the Martian atmospheric emission in the dayside near-terminator region during varying solar flare intensities.

## 7. Data Availability

The MAVEN data utilized during this work are accessible through the NASA Planetary Data System at https://pds-ppi.igpp.ucla.edu/search/?t=Mars&sc=MAVEN&facet=SPACECRACT _ NAME&depth=1. The MAVEN IUVS Level 1c, version_13, revision_01 data product bundles (Schneider et al., 2015), SWEA Level 2, version_17, revision_07, in-situ calibrated data bundles (Dunn et al., 2023; Mitchell et al., 2015), and EUV Monitor Levels 2 & 3, version_15, revision_01 calibrated and modelled datasets, respectively (Eparvier, 2023) were used during this research work and SWEA data can also be accessed through the Python Data Analysis and Visualization tool (MAVEN SDC et al., 2020). The MAVEN SDC is utilized to identify solar flares at Mars and can be accessed using this link: https://lasp.colorado.edu/maven/sdc/public/pages/notebook/events/index.html#/.

## 8. Acknowledgements

We sincerely acknowledge the NASA PDS and MAVEN team especially SWEA and EUV for the datasets. L. Ram and R. Sharma acknowledge the fellowship from the Ministry of Education, Government of India for carrying out this research work. R. Rathi acknowledges the fellowship from the Innovation in Science Pursuit for Inspired Research (INSPIRE) programme, Department of Science and Technology (DST), Government of India. The support from Sonal K. Jain and Nick Schneider is duly acknowledged. This work is also supported by the Department of Space and the Ministry of Education, Government of India.